\documentclass[aps,prl,twocolumn,groupedaddress,showpacs]{revtex4}

\usepackage{graphicx}
\begin{document}

\title{Directed motion for delta-kicked atoms with broken symmetries: comparison between theory and experiment}

\author{P. H. Jones, M. Goonasekera,  D. R. Meacher\\T. Jonckheere \& T. S. Monteiro}
\email[]{philip.jones@ucl.ac.uk}

\affiliation{Department of Physics and Astronomy, University
College London, Gower Street, London, United Kingdom, WC1E 6BT}

\date{\today}

\begin{abstract}
We report an experimental investigation of momentum diffusion in the
$\delta$-function kicked rotor where time symmetry is broken by a
two-period kicking cycle and spatial symmetry by an alternating linear 
potential. We exploit this, and a technique involving a moving
optical potential, to create an asymmetry in the momentum
diffusion that is due to the classical chaotic diffusion. This represents
a realization of a type of Hamiltonian quantum ratchet.
\end{abstract}


\maketitle

The ratchet effect, in other words the rectification of noise in a system
without net bias, was first proposed by Feynmann and has since formed the subject of 
numerous studies \cite{Reimann}.
Recently, there has been further interest and investigations of Hamiltonian 
chaotic ratchets, where the extrinsic noise is replaced by deterministic chaos.
Hamiltonian systems offer the additional possibility of a fully quantum ratchet,
where some form of directed transport appears in the context of coherent wave
dynamics; other types of ratchets in dissipative and noisy quantum
systems, corresponding to coherence times which are relatively short,
 have also been proposed \cite{Hanggi2}. 
 Previous studies of chaotic Hamiltonian ratchets \cite {Mixed} indicated that directed motion arises if
certain symmetries are broken, but persists only in the presence of mixed phase-space
dynamics (eg a bounded classical phase-space with a mixture of regular tori and chaotic regions).
 The quantitative analysis of the directed current then relies  on the details 
of the classical phase-space.

In \cite{Monteiro2002, Jonckheere} an alternative theoretical 
proposal for chaotic, but asymmetric momentum diffusion, aimed at a realization
 with cold atoms in far-detuned pulsed optical lattices, was presented. 
Experiments with cold atoms in {\em near-detuned}, driven  optical lattices,
had already been shown to provide realizations of classical 
Brownian and dissipative ratchets 
\cite{Schiavoni}. {\em Far-detuned} lattices minimize decoherent effects: hence they
 provided the clearest
demonstrations of Hamiltonian quantum chaotic dynamics\cite{Raizen}.
In particular, cold atoms in $\delta$-kicked optical lattices can realize the dynamics of
the chaotic quantum kicked rotor (QKR) and show the effect of dynamical
localization (DL): the momentum diffusion of the cold atoms follows approximately
the classical chaotic rate, $\langle p^2\rangle  \approx Dt$,
 up to a timescale $ t^* \propto \hbar^{-2}$, after which the
diffusion stops and the quantum momentum probability distribution $N(p)$ localizes, 
with a variance $ \langle p^2 \rangle^{1/2} =  L \sim D/\hbar$. DL is a quantum coherent effect due to destructive
wave interference \cite{Fish}.

 The directed chaotic transport mechanism of \cite{Monteiro2002,Jonckheere}
 is generic in character: a quantum kicked rotor with broken time and space symmetry 
diffuses asymmetrically for a finite timescale $t_R$. For the classical system, unfortunately
$L$ grows without limit as $t \to \infty$. For the quantum dynamics, localization
 arrests the diffusion, and a momentum asymmetry  which is comparable to $L$ can result.
 This may be considered a type of {\em quantum} ratchet. An early proof-of-principle
experimental realization was carried out \cite{preprint}. Other 
theoretical proposals for Hamiltonian regular or chaotic directed motion
have subsequently also been published by several groups \cite{Hamiltonian}.

In this paper we report  experimental realizations of the kicked rotor 
with broken time and space symmetry which  permit for the first time
quantitative comparisons with the 
analytical results for the `ratchet time' $t_R$ and the periodic `current 
 reversals' expected from the theoretical model \cite{Monteiro2002,Jonckheere}.
We present an accurate formula for the classical ratchet current (a simplified
formula was presented in \cite{Hur}). 
Good agreement is obtained between theory and experiment. To our knowledge,
this remains the only experimental realization of directed transport in a 
Hamiltonian quantum system.

An optical lattice formed by two counter-propagating laser beams
may be used to trap laser-cooled atoms in a one-dimensional
periodic potential. Here, an accelerating optical lattice was sometimes also used 
to apply an additional `rocking' linear
potential. In a frame with acceleration $a$,
 the Hamiltonian has an additional inertial term \cite{Madison1999}:
\begin{equation}
H = \frac{p^2}{2M} + V_0\cos(2k_L x) \pm Ma x
\label{Hamiltonian}
\end{equation}
where $M$ is the mass of the atom, $k_L = 2\pi / \lambda$ the
laser wavevector and $V_0$ the potential depth. If the optical
lattice is applied as a series of short ($\delta$-function) pulses
with period $T$, then we may as for the usual
 $\delta$-kicked rotor, write the Hamiltonian including 
the rocking potential in dimensionless
form:
\begin{equation}
\mathcal{H} = \frac{\rho^2}{2} + \sum_n \left[K\cos(\phi)+ A (-1)^n \phi \right] \ \delta(\tau - n)
\label{kick}
\end{equation}
where $K$ is the stochasticity parameter which describes the
strength of the kick. Here $\rho = 2Tk_Lp/M$ is a scaled
momentum, $\phi = 2k_Lx$ a scaled position, $\tau = t/T$ a scaled
time and $\mathcal{H} = 8\omega_RT^2H/\hbar$ the scaled
Hamiltonian. The commutation relation $[\phi,\rho] = i8\omega_RT$
gives the scaled unit of system action or effective Planck
constant $\hbar_{ef\!f} = 8\omega_RT$ ($\omega_R$ the atomic recoil
frequency, $2\pi \times 2.1$~kHz for cesium D2) which may be controlled  through the period of the
pulses.

For a $\delta$-kicked rotor, to lowest order, the
momentum diffusion rate is the uncorrelated rate $D \simeq K^2/2$ . Corrections to this arise
from correlations between kicks  \cite{Rechester1980}.
In \cite{Jonckheere} it was shown that for the classical $\delta$-kicked
rotor where time symmetry is broken by a two-period kicking cycle
of periods $T(1+b):T(1-b)$ (where $b \ll 1$), then, for short times, 
the correlations produce an asymmetry in the momentum distribution $N(\rho)$. 

Here, we considered an ensemble of particles with initial ($t=0$) momentum
distribution strongly peaked about a value $\rho_L$, in the lattice frame, ie
$N_0(\rho)\approx \delta(\rho-\rho_L)$. Using the method of \cite{Rechester1980}, we
calculated a value for the asymmetry $I(t) = \langle  (\rho - \rho_L)\rangle $ at later times:

\begin{eqnarray}
I(t)= I_0 \sin \left( (1-b)A- 2b \rho_L \right) F(t)
\label{current}
\end{eqnarray}
where the maximum current is \\
$I_0=\frac{-K \ J_1(2Kb)}{1-J_0(2Kb)^2}
\left[J_0(2Kb)J_2((1-b)K)+J_2((1+b)K) \right]$.
and the time-dependence is given by  $F$ where $1 \geq F(t) \geq 0$:
\begin{eqnarray}
F(t)= {1-J_0(2Kb)^{2t-2}}.
\end{eqnarray}
For $t$ small, $F(t)\sim t$ grows linearly with time, but for $t \gg t_R \approx 1/(Kb)^2$,
it saturates, ie $F(t) \to 1$. Hence $t_R$ is the classical timescale
for the current to develop; if dynamical localization  occurs too quickly, ie
 $t* \ll t_R$, no appreciable quantum effect is observed. Conversely if
$t^* \gg t_R$, the asymmetry is negligible compared with $L$,
(the asymptotic  variance of $N(\rho)$). Optimally, we require $t^* \sim t_R$.

For a ratchet effect, as usually understood, a distribution initially
with zero average momentum $\langle \rho\rangle =0$ , and $\rho_L=0$)
ie in the rest frame of a potential (giving
no net bias), evolves to an asymptotic distribution with
 $\langle \rho\rangle  \geq 0$. In the present experiment, this
requires $A \neq 0$: for example, for $A=\pi/2$, and $\rho =0$, if at $t=0$,
$\langle \rho\rangle =0$, we obtain $I(t >  t^*) \simeq I_0$. However, in order to 
fully investigate the underlying mechanism, here we also investigated extensively
starting conditions with non-zero initial momentum, ie $\rho \neq 0$.
  
\begin{figure}
\includegraphics[width=0.6\columnwidth]{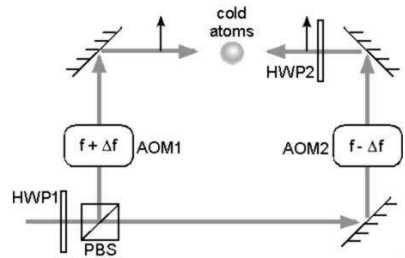}
\caption{Diagram of apparatus\label{ApparatusDiagram}. The
half-wave plate HWP1 and polarizing beam splitter PBS are used to
create two equal intensity beams which are shifted in frequency by
$f\pm \Delta f$ by the AOMs. HWP2 is used to make the
polarizations of the two beams parallel.}
\end{figure}

In our experiment we use laser-cooled cesium atoms in a far-off resonant pulsed
optical lattice. The lattice is formed by two horizontal
counter-propagating laser beams, $1/e$ radius (0.95$\pm$0.05~{\rm mm}),
with parallel linear polarizations (see
figure~\ref{ApparatusDiagram}) which produces a spatial variation
of the AC Stark shift proportional to the local intensity,
and which hence, is sinusoidal.

The pulses are produced by rapidly switching the drive voltage to
the acousto-optic modulators (AOMs) according to a pre-defined
sequence. The time between the kicks may be altered in order to
produce the two-period $T(1+b):T(1-b)$ alternating kicking cycle described above.
The experiment proceeds as follows. Cesium atoms are trapped and
cooled in a standard six-beam magneto-optic trap (MOT) before
further cooling in an optical molasses to an rms scaled momentum
width of $\sigma_{\rho} \simeq 4$. The molasses light is turned
off using an AOM and the periodic ``kicking" potential applied.
The beams for the kicking potential are derived from a Ti:Sapphire
laser with an output power of 1~W at 852~nm, detuned typically 2000
linewidths (natural linewidth $\Gamma$ = 2$\pi \times$ 5.22~MHz) to
the low frequency side of the D2 cooling transition in cesium, which is
sufficient for effects due to spontaneous emission to be
neglected. This is split into two equal intensity beams using a
half-wave plate and polarizing beam splitter (HWP1 and PBS in
figure~\ref{ApparatusDiagram}) and each beam sent through an AOM.
The two AOMs are driven by separate (phase-locked) radio-frequency
synthesizers that are controlled by separate fast
radio-frequency switches but triggered by the same arbitrary
function generator that produces the kicks. After the kicking the
cloud of cold atoms is allowed to expand ballistically for up to 20~ms 
before a pair of counter-propagating near-resonant
laser beams are switched on and the fluorescence from the atoms
imaged on a CCD camera. From the spatial distribution of the
fluorescence it is then possible to extract the momentum
distribution. Using this apparatus we have checked that
(for regularly spaced kicks, i.e. $b=0$ in the above) 
dynamical localization 
can be observed  as a change to an exponential momentum distribution
for $t^* \sim K^2/\hbar_{eff}^2$.

To investigate starting conditions with non-zero mean momentum we have used a moving optical lattice formed by laser beams with a controlled frequency difference to make the kicking potential, so that atoms which are stationary in the laboratory frame have a momentum $\rho_L$ in the rest frame of the optical potential.  This is achieved by driving the AOMs at frequencies that differ by $2\Delta f$, such that the atomic momentum in the rest frame of lattice is $\rho_L = m\lambda^2 \Delta f \hbar_{ef\!f}/4\pi \hbar$.  Using this technique, $\rho_L$ may be varied over a
large range in order to sample several periods of the oscillation of the asymmetric diffusion without the beams becoming significantly misaligned from the cloud of cold atoms.

\begin{figure}
\includegraphics[height=2.0in]{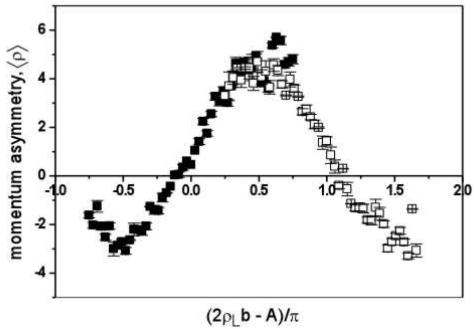}
\caption{Demonstration of a type of quantum `ratchet effect'.
The graph shows the final momentum asymmetry $\langle \rho \rangle$ 
vs $\Phi = (2\rho_L b-A)/\pi$ for $K$ = 2.6, $b$=1/16, $\hbar_{ef\!f}=1$.
 Filled squares are data for $\rho_L = 0$, open squares
are $\rho_L = 8\pi$. For the maximum at $\Phi=0.5$, the cloud is initially
at rest relative to the optical lattice (ie $\rho_L =0$) and
has no asymmetry; after 120 kicks, there is a constant `ratchet current' $\langle \rho\rangle \approx 4$.}
\label{Fig2}
\end{figure}

\begin{figure}
\includegraphics[height=2.5in]{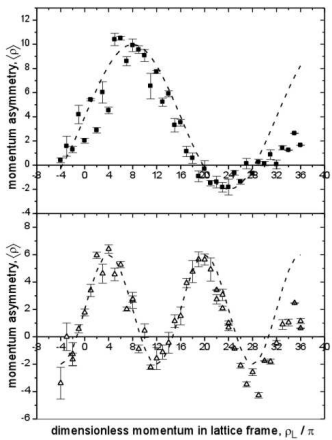}
\caption{ Behavior as a function of $b$ and initial momentum $\rho_L$.
Momentum asymmetry \textit{vs} starting momentum $\rho_L$ in the
lattice frame for $K=3.3$, $\hbar_{eff}=1$, $b=1/32$ (filled
squares) and $b=1/16$ (open triangles). The asymmetry is 
consistent with the theoretical prediction of Eq.\ref{current}, ie
 $\langle \rho \rangle \propto I_0 \sin 2\rho_Lb$ 
and (approximately) $I_0 \propto 1/b$.\label{Fig3}}
\end{figure}

For these experiments the period of the kicks is $T = 9.47~\mu{\rm s}$
and pulses are square with duration typically $t_p = 296~{\rm ns}$
($t_p/T = 1/32\leq b$), which is sufficient for there to be no substantial
 effects on the diffusion constant due to the finite temporal width of the kicks in the region
 of $\rho_L \approx 0$ \cite{Klappauf1999} (for larger $\rho_L$ these effects become 
important and start to affect the data). An investigation of the effects
on the momentum diffusion arising the finite width of the kicks
 was presented in \cite{Jones2004c}.  
We were able to investigate values of the parameters $K =2 \to 5$ 
(with order 10\% error arising mainly from the measurement of the beam intensity),
$\hbar_{ef\!f} = 1$, and values of $b=1/8 \to 1/32$. 
 Values of  $K$ close  to the first maximum 
of the Bessel function $J_2(K)$  may be expected to produce the largest maximal
currents $I_0$ and hence the clearest experimental signature. For
values of $K \approx  2-5$ there are still stable islands in the classical phase-space;
nevertheless these are sufficiently small:  for $b \neq 0$, good quantitative 
agreement with calculated chaotic
diffusive rates is obtained for lower $K$ than for the Standard Map \cite{Jonckheere}.

Finally, for experiments with $A \neq 0$, the linear `rocking' term of alternating 
sign was included by 
accelerating the optical lattice. This is done by modulating 
the frequency of one of the laser beams in a linear manner using a second 
(phase-locked) arbitrary function generator by an amount $\pm\delta\! f$ in the 
time of the kick period $T$. 
The dimensionless potential gradient $A$ is related to the magnitude of the 
frequency modulation (acceleration of the lattice) by $A = 2\pi t_p\delta\! f$
 for finite square pulses of width $t_p$. Accelerating the potential thus provides a 
simple way of controlling the magnitude of $A$ and hence controlling the phase 
shift of the momentum-dependent diffusion constant in order to make it locally 
asymmetric around zero momentum. As the maximum frequency modulation amplitude allowed by 
the radio-frequency synthesizers was $\pm 1.25~{\rm MHz}$ this limits the range of $A$ 
achievable to $\pm3\pi /4$. In order to observe one complete oscillation of the momentum
 diffusion constant, for some experiments an additional constant frequency offset was
 introduced between the laser beams such that in the rest frame of the lattice the mean
 atomic momentum was $\rho_L = 8\pi$.

Figure~\ref{Fig2} shows the asymmetry $\langle \rho \rangle$
 (combining $\rho_L = 0$ and $\rho_L = 8\pi$) plotted as a function of 
$\Phi= (2\rho_Lb-A)/\pi \approx  (2\rho_Lb-A(1-b))/\pi$ suggested by  Eq.\ref{current}.
 In particular, the data at $\Phi=0.5$,
corresponding to $\rho_L=0$ hence represents a ratchet current $I_0 \approx 4$
obtained in the rest frame of the potential. The behavior is in good agreement
 with quantum simulations carried out here, which yielded
 $I(t \geq t^*) \simeq I_0 \sin \Phi \simeq 3.8 \sin \Phi$.
The corresponding classical formula from Eq.\ref{current} yields a much larger
current $I_0 \approx 7.5$. For these values of $K,b$ the result is 
close to the simplified classical current formula 
$I(t \to \infty) \approx K^2 \frac{J_2(K)}{b} \sin \Phi$ 
obtained in \cite{Hur}. 

\begin{figure}
\includegraphics[width=1.0\columnwidth]{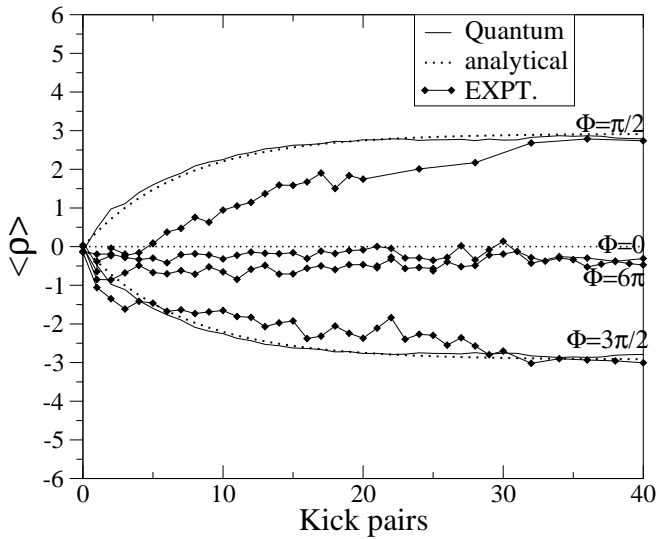}
\caption{Time dependence of the momentum asymmetry 
for $K=2.1$, $\hbar_{eff}=1/4$, and $b=1/8$ (filled squares), 
for experimental, classical and quantal results in a regime where
$t^* \simeq t_R$. Good agreement for all three is shown. 
The results confirm the validity of the time-dependence of
the formula Eq.\ref{current}.
 \label{Fig4}}
\end{figure}

\begin{figure}
\includegraphics[height=2.5in]{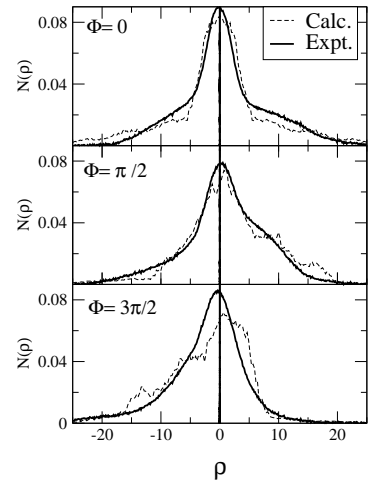}
\caption{Experimental and quantal
momentum distributions after dynamical localization, corresponding to Fig4. }
\label{Fig5}
\end{figure}

In Fig.\ref{Fig3} the dependence on  $\langle \rho \rangle$ on $b$  is tested more extensively
for $A=0$ and is seen to oscillate with a period 
$\pi/b$ and is consistent with $I_0 \propto 1/b$, in agreement with theory.
However, in order to test experimentally the time dependence of the classical correlations
(ie the behavior of $F(t)$ above), further experiments for smaller $\hbar \approx 1/4$,
for which there is closer agreement for classical and quantal diffusion (ie $t^* \approx t_R$),
were obtained. Hence in Fig.\ref{Fig4}, a comparison of the behaviour of the
current $\langle \rho \rangle$ is undertaken for
quantal, classical (ie Eq.(\ref{current}) and experimental values and for $K=2.1$, $\hbar=1.4$.
The agreement between all three is quite good. Strikingly, there is for
this value of $\hbar$, near-perfect agreement between the quantum and classical
time dependence. For these parameters, as investigated in \cite{Jonckheere},
the momentum diffusion has minima (ie $D(p_0)$ is rather small for $ p_0 \approx \pi/(4b)$.
The quantum cloud localizes as soon as  the variance $L \sim \pi/(4b)$, hence $t^* \sim \pi^2/(Kb)^2 \sim t_R$.
For the classical cloud, diffusion continues (albeit more slowly) and the momentum
variance increases with time but the asymmetry saturates.
Hence, the close agreement between the classical and quantal {\em asymmetry} is not unexpected.
Fig.\ref{Fig5} shows the corresponding calculated quantal and experimental momentum
distributions after dynamical localization (120 kicks) for the same parameters as
Fig.\ref{Fig4}. We note that the method for evaluation of the experimental
$N(\rho)$ will tend to somewhat over-estimate $<\rho>$; conversely, the theoretical results
are for $\sigma_p=1$; for larger initial momentum widths, there will be a damping of the
current $I_0 \to I_0 \exp{-4\sigma_p^2 b^2}$. Although the near exact quantitative agreement at $t\to \infty$
is somewhat fortuitous, theory and experiment are in quite good agreement. 

In sum, we have demonstrated experimentally  directed transport
which relies on the persistence of coherent, unitary quantum evolution over
the full timescale of the experiment. This represents a type of Hamiltonian quantum
ratchet.
 
\begin{acknowledgments}
We would like to thank H.E.Saunders Singer for useful discussions and EPSRC and UCL for financial
support.
\end{acknowledgments}

\end{document}